\documentclass[a4paper,11pt]{article}
\pdfoutput=1 

\usepackage{jinstpub} 
\usepackage{caption}
\usepackage{subcaption}
\usepackage[official]{eurosym}
\usepackage{graphicx}
\usepackage{hyperref}

\title{\boldmath Light-Trap: A SiPM Upgrade for VHE Astronomy and  Beyond}

\author[a,1]{J. E. Ward,\note{Corresponding author.}}
\author[a]{J. Cortina,}
\author[a]{D. Guberman}

\affiliation[a]{Institut de Fisica d'Altes Energies (IFAE), The Barcelona Institute of Science and Technology, \\Campus UAB, 08193 Bellaterra (Barcelona), Spain}

\emailAdd{jward@ifae.es}

\abstract{Ground-based gamma-ray astronomy in the Very High Energy (VHE, E$>$100 GeV) regime has fast become one of the most interesting and productive sub-fields of astrophysics today. Utilizing the Imaging Atmospheric Cherenkov Technique (IACT) to reconstruct the energy and direction of incoming gamma-ray photons from the universe, several source-classes have been revealed by previous and current generations of IACT telescopes (e.g. Whipple, MAGIC, HESS and VERITAS). The next generation pointing IACT experiment, the Cherenkov Telescope Array (CTA), will provide increased sensitivity across a wider energy range and with better angular resolution.

With the development of CTA, the future of IACT pointing arrays is being directed towards having more and more telescopes (and hence cameras), and therefore the need to develop low-cost pixels with acceptable light-collection efficiency is clear.

One of the primary paths to the above goal is to replace Photomultiplier Tubes (PMTs) with Silicon-PMs (SiPMs) as the pixels in IACT telescope cameras. However SiPMs are not yet mature enough to replace PMTs for several reasons: sensitivity to unwanted longer wavelengths while lacking sensitivity at short wavelengths, small physical area, high cost, optical cross-talk and dark rates.

Here we propose a novel method to build relatively low-cost SiPM-based pixels utilising a disk of wavelength-shifting material, which overcomes some of these drawbacks by collecting light over a larger area than standard SiPMs and improving sensitivity to shorter wavelengths while reducing background. We aim to optimise the design of such pixels, integrating them into an actual 7-pixel cluster which will be inserted into a MAGIC camera and tested during real observations.

Results of simulations, laboratory measurements and the current status of the cluster design and development will be presented.}


\begin{document}
\maketitle
\flushbottom

\section{Introduction}
\label{sec:intro}

VHE gamma rays are detected using IACT telescopes (optical reflectors and fast electronics) by measuring very short (1-2 ns) Cherenkov light flashes produced by gamma-ray initiated extensive air showers (EAS) in the upper atmosphere. The detection of these EAS is challenging because of their extremely short duration and low light intensity, as well as the presence of vastly greater number of EAS produced by background hadrons and electron/positrons entering the atmosphere. 

Some of the future physics goals in VHE astronomy will necessitate the ability of IACT telescope systems to observe several-degree sections of the sky at one time \cite{CTA, machete}, thus allowing efficient large-scale studies like sky surveys for unbiased cataloguing and characterisation of sources of VHE emission, e.g. Active Galactic Nuclei or Supernova Remnants. Observing large portions of the sky also increases the likelihood of detecting Gamma-Ray Bursts (GRBs) serendipitously in the FOV, long since a goal of ground-based gamma-ray astronomy. Furthermore, larger FOVs will allow for detailed morphological studies of galactic sources with relatively large angular sizes and the diffuse gamma-ray emission along the galactic plane.

\begin{figure}[htbp]
   \centering
   \includegraphics[width=0.6\textwidth]{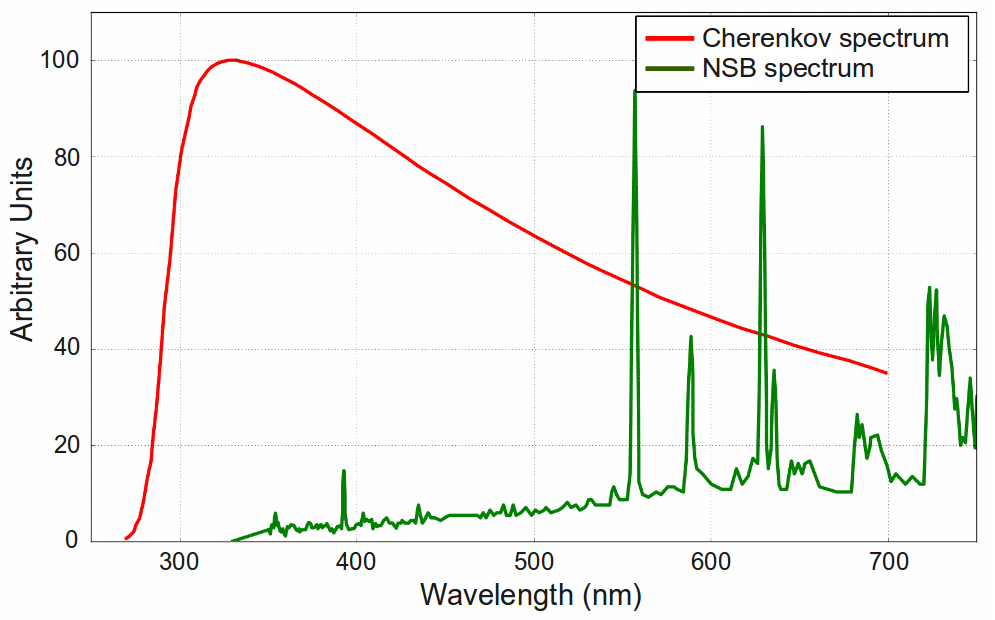} 
   \caption{The Night Sky Background (NSB) \cite{a} and Cherenkov spectrum at 2200 meter altitude in La Palma, Spain \cite{b}.}
   \label{fig:NSB}
\end{figure}

To achieve these goals, a larger camera with an expanded FOV would need to be constructed. However, the cost of making larger and larger cameras needing many thousands of pixels, if utilising PMTs at $>$ \euro{}100/pixel, rapidly becomes prohibitive. Replacing PMTs with SiPMs is the most effective approach to solving this dilemma and is hence an active field of research and development in the VHE astronomical community. SiPMs may soon become much cheaper than PMTs, provide 50$\%$-100$\%$ higher photo-detection efficiency (PDE), can be calibrated easily and do not operate under high voltage. However SiPMs are currently not perfectly suited for VHE astronomy, as individual units are not commercially available in sizes larger than 10$\times$10 mm$^{2}$, they still cost $>$1 \euro{}/mm$^{2}$, reach their PDE at a wavelength longer than the peak emission wavelength of Cherenkov radiation in EAS (see Figure \ref{fig:NSB}), and have too high a PDE at longer wavelengths where there is much less signal from EAS but a large background contribution from the night sky.

\section{The Light-Trap}

\begin{figure}[htp]
  \centering
  \subcaptionbox{Top down view\label{fig:LTTop}}{\includegraphics[width=0.50\textwidth]{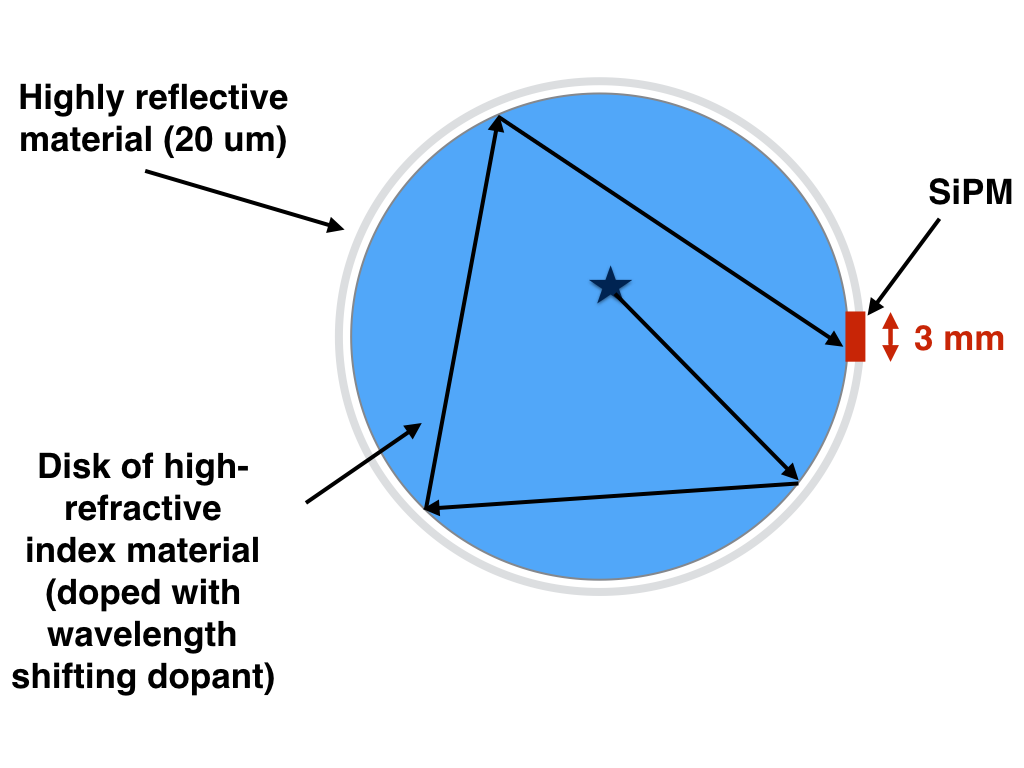}}\hspace{-0.0em}%
  \subcaptionbox{Side view\label{fig:LTSide}}{\includegraphics[width=0.50\textwidth]{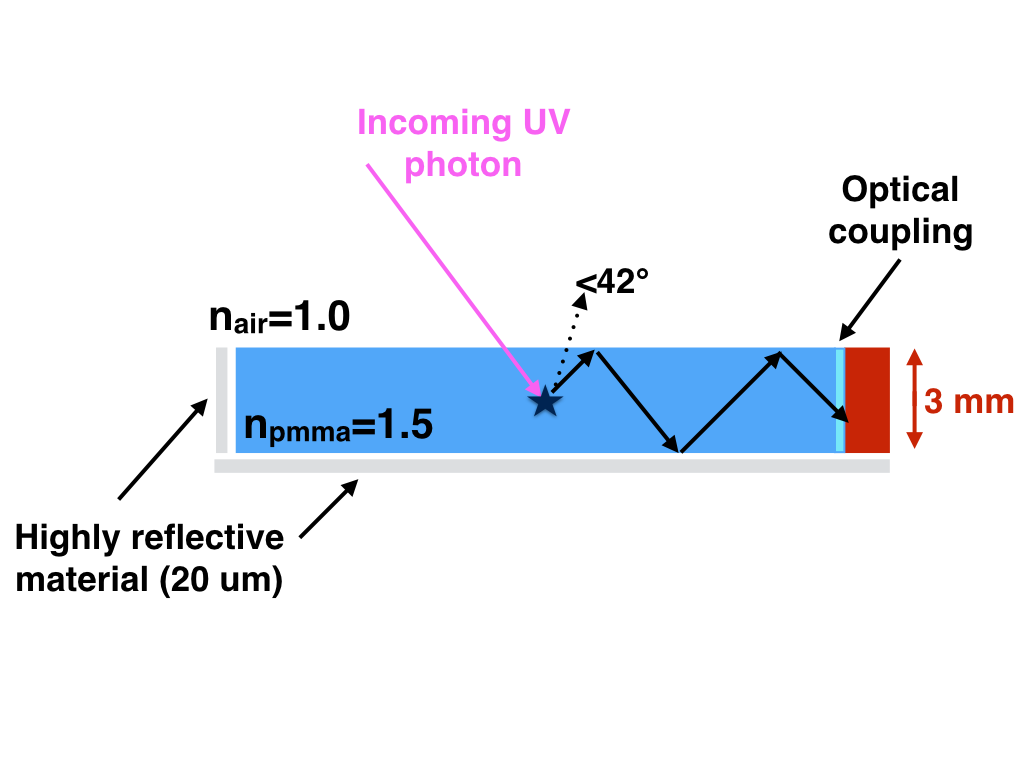}}
  \caption{Conceptual design of the Light-Trap.}
  \label{fig:LTconcept}
\end{figure}

Considering the previously stated disadvantages to the use of SiPMs in VHE astronomy, we propose to solve most of these issues by attaching a SiPM device to a "Light-Trap" disk (see Figure \ref{fig:LTconcept}). This disk will contain some wavelength-shifting (WLS) fluors which absorbs photons in the \mbox{$\sim$300-400 nm} wavelength range and re-emits them in the $\sim$400-500 nm range. WLS photons are re-emitted isotropically, so a fraction of them (approximately 75\% for a material with n=1.5 refractive index) are trapped in the disk by total internal reflection (TIR) and eventually reach the SiPM. The rest of the re-emitted photons escape.

As a result,

\begin{enumerate}
  \item Light around the peak of the Cherenkov spectrum ($\sim$350 nm) is collected.
  \item Light at longer wavelengths (for which the NSB dominates) is
rejected.
  \item The PDE of the detector effectively shifts by about 100 nm relative to the PDE
of the SiPM so that its peak gets nearer to the peak of the Cherenkov spectrum.
  \item The collection area of the detector can be a factor $\sim$10-50 larger than the sensitive area of the SiPM, i.e. the cost is reduced by the same factor (if the cost of the disk is low) and thus enables us to build pixels far larger than commercially available SiPMs.
\end{enumerate}

\section{Proof-of-Concept Laboratory Measurements}

\begin{figure}[htp]
  \centering
  \subcaptionbox{EJ-299-15 absorption and emission spectra\label{fig:EJ29915}}{\includegraphics[width=8cm]{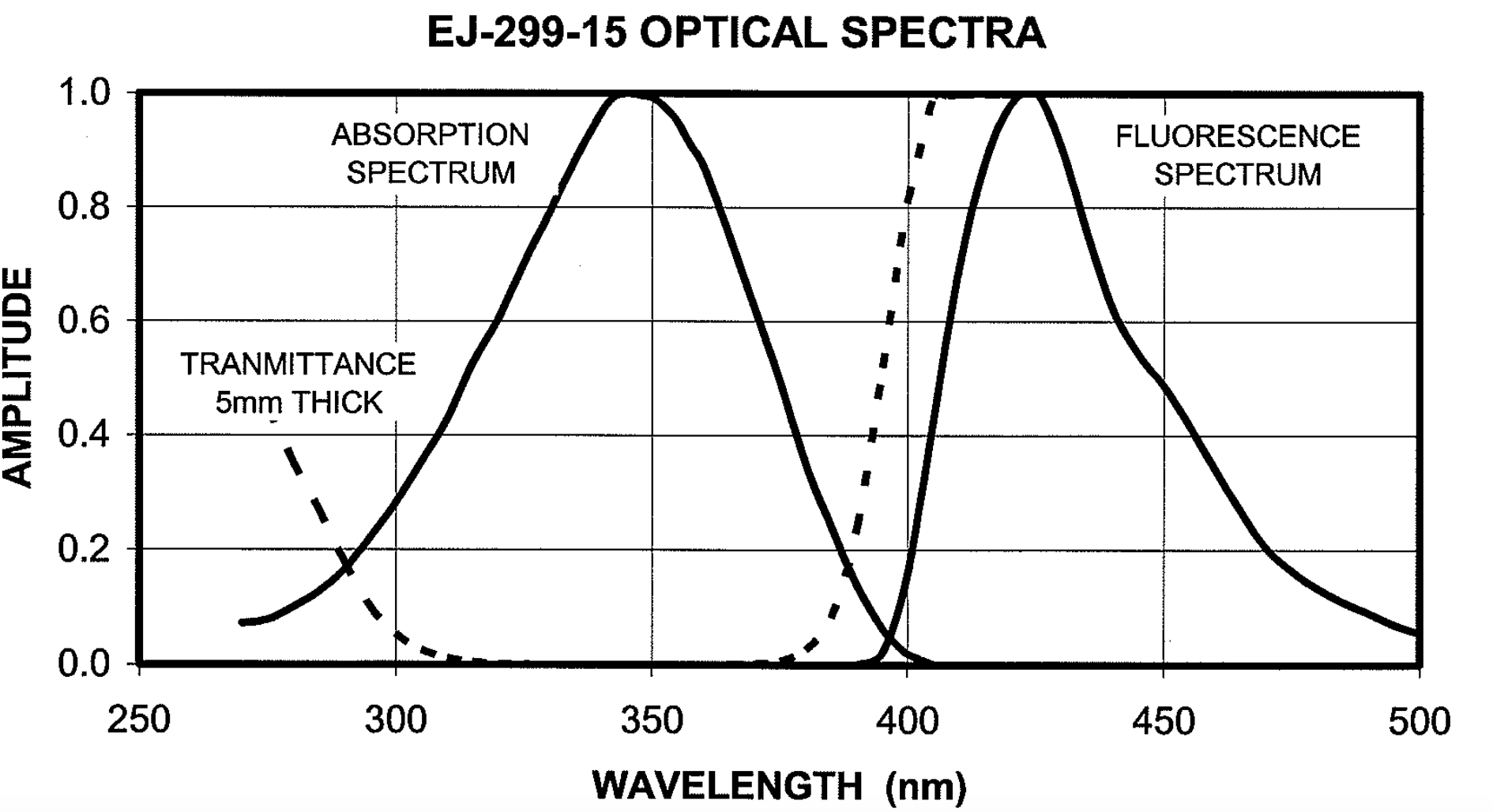}}\hspace{1em}%
  \subcaptionbox{DSB-1 absorption and emission spectra\label{fig:DSB1}}{\includegraphics[width=8cm]{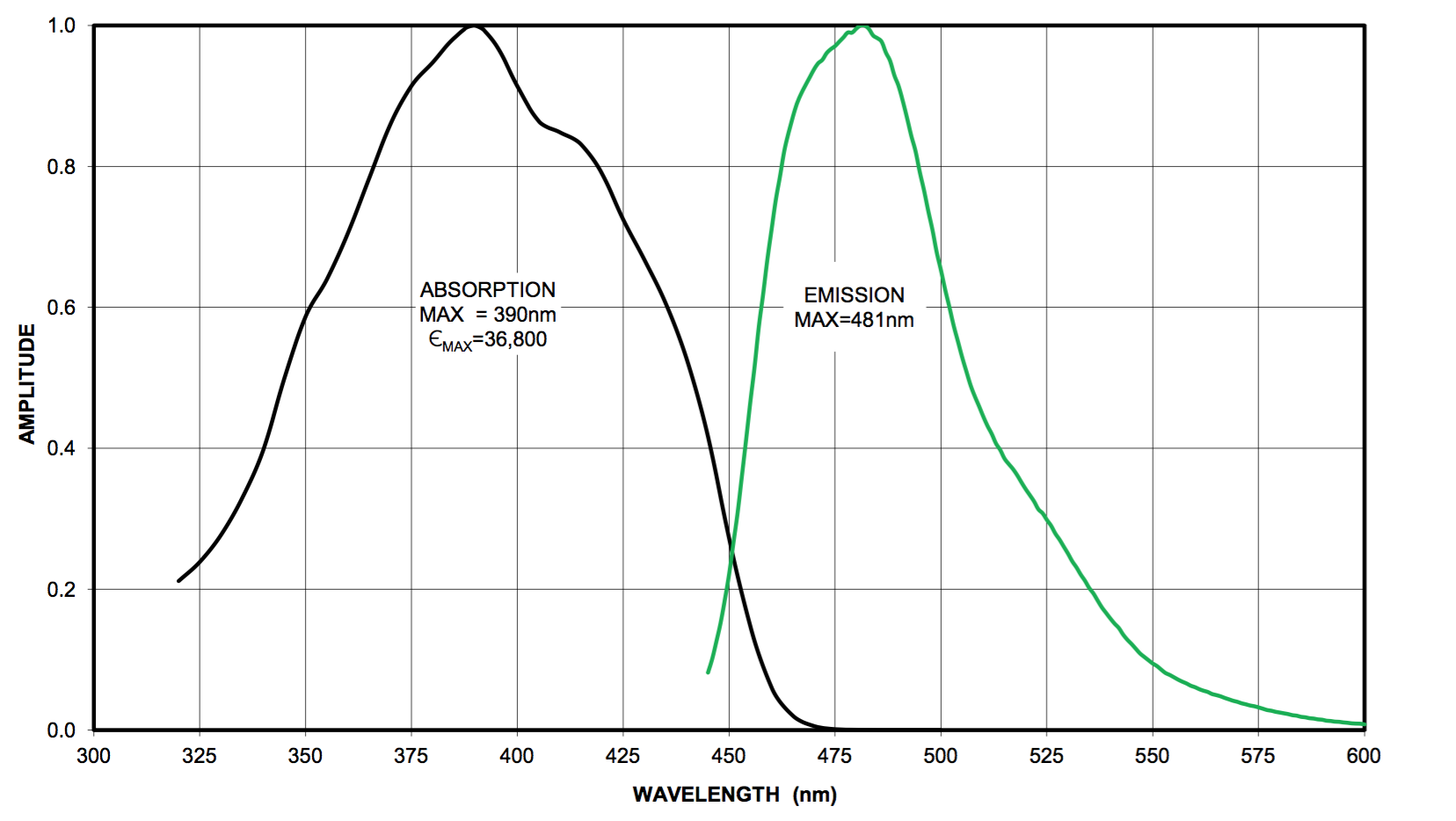}}
  \caption{Spectra of the two Eljen wavelength-shifting materials, as provided by the company.}
\end{figure}

To produce a proof-of-concept device, custom-doped wavelength-shifting polymethylmethacrylate (PMMA, refractive index n=1.49) disks were purchased from Eljen Technology (Sweetwater, Texas). Two different sets of disks of 15mm diameter with a 3mm thickness were purchased. One set (material designation EJ-299-15), is intended to absorb in the UV and re-emit in blue-violet (see Figure \ref{fig:EJ29915}) while the second set was doped with the fluor DSB1 which is also sensitive to UV but re-emits in green (see Figure \ref{fig:DSB1}). The dopant levels were customised by Eljen according to our specifications i.e. to absorb 100\% of incident 340 nm photons within 1.5 mm of the material. The wavelength-shifting fluors added to the disks also have fast responses on the order of $\sim$3 ns, with a quantum yield of $\sim$84\%. 

The SiPM used for this study was the 3mm x 3mm KETEK PM3375 (with a peak sensitivity at 420 nm), which was coupled to the Eljen disk using an optically clear silicone rubber sheet (EJ-560, also purchased from Eljen Technology) that was cut to the SiPM size accordingly. This silicone material is soft and only lightly adhesive, thus allowing the removal and addition of the SiPM. To help improve efficiency in the event that wavelength-shifted photons are not total internal reflected or escape due to optical imperfections, 3M\textsuperscript{\textregistered} reflective foil was cut so as to surround the back and sides of the disk. 

\begin{figure}[htbp]
   \centering
   \includegraphics[width=0.4\textwidth]{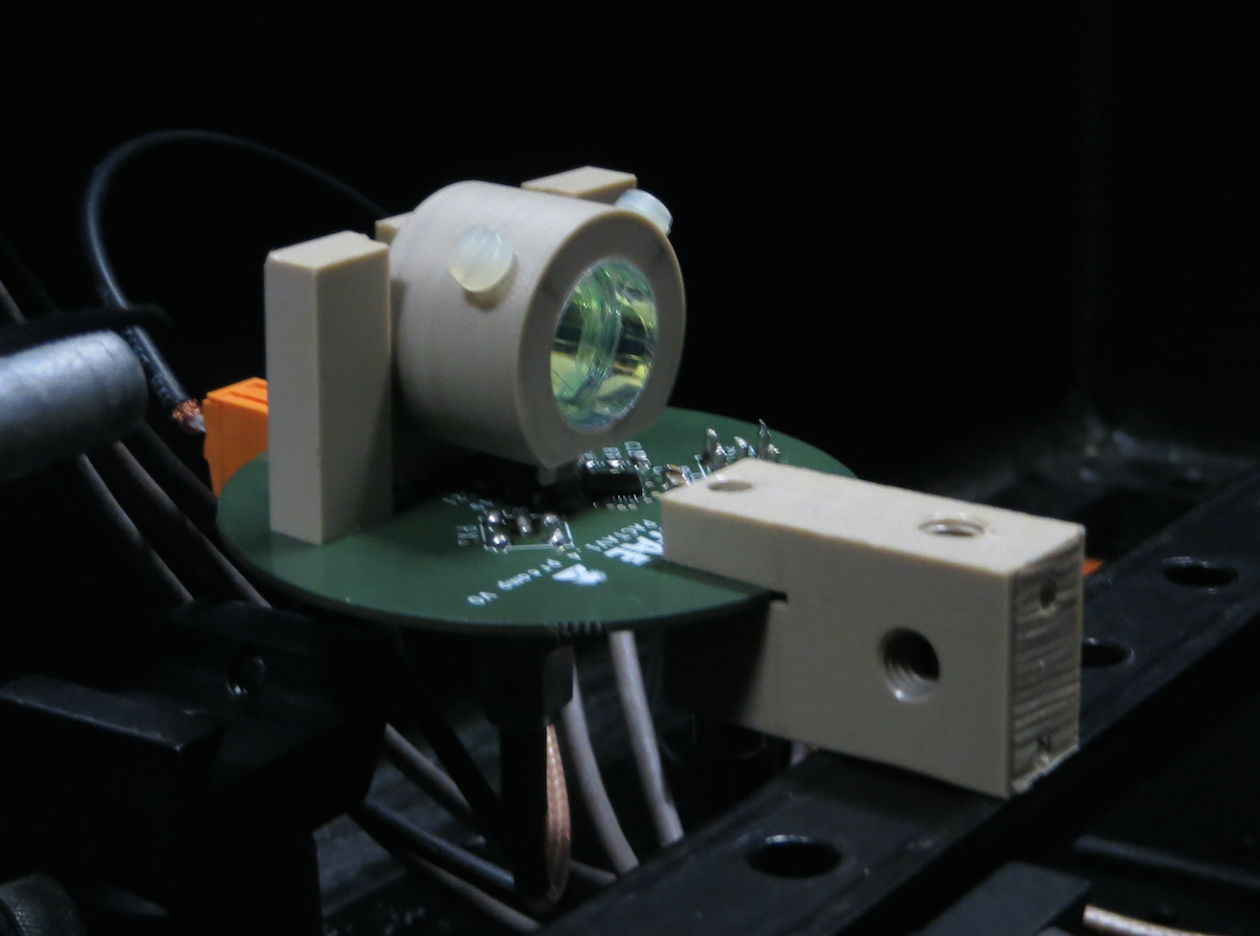} 
   \caption{Image of the Light-Trap holder and electronics board. The SiPM is facing upwards in this picture and the disk surface is pointing to the right.}
   \label{fig:LTHold}
\end{figure}

In order to hold the SiPM, PMMA disk, and reflective foil together it was necessary to construct a cylindrical polyethylene holder (see Figure \ref{fig:LTHold}). This holder could be mounted on the SiPM pre-amp and power board, had screws to keep the detector elements secured in place, and also had a window to allow a SiPM to attach to the side of the disk. It should be noted that the reflective foil was firmly positioned inside this holder, although it is still expected to be in contact with the disk in places (thus impacting TIR efficiency). 

\begin{figure}[htp]
  \centering
  \subcaptionbox{Laboratory set-up.\label{fig:scheme}}{\includegraphics[width=7cm]{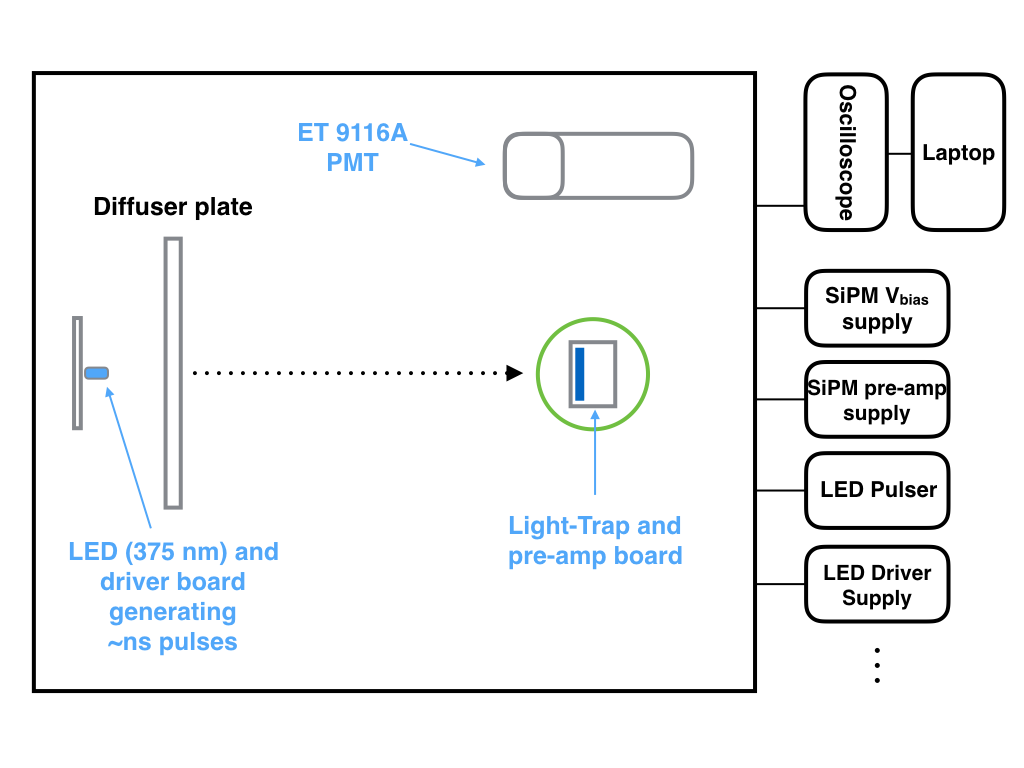}}\hspace{1em}%
  \subcaptionbox{Light sensor configurations.\label{fig:orient}}{\includegraphics[width=7cm]{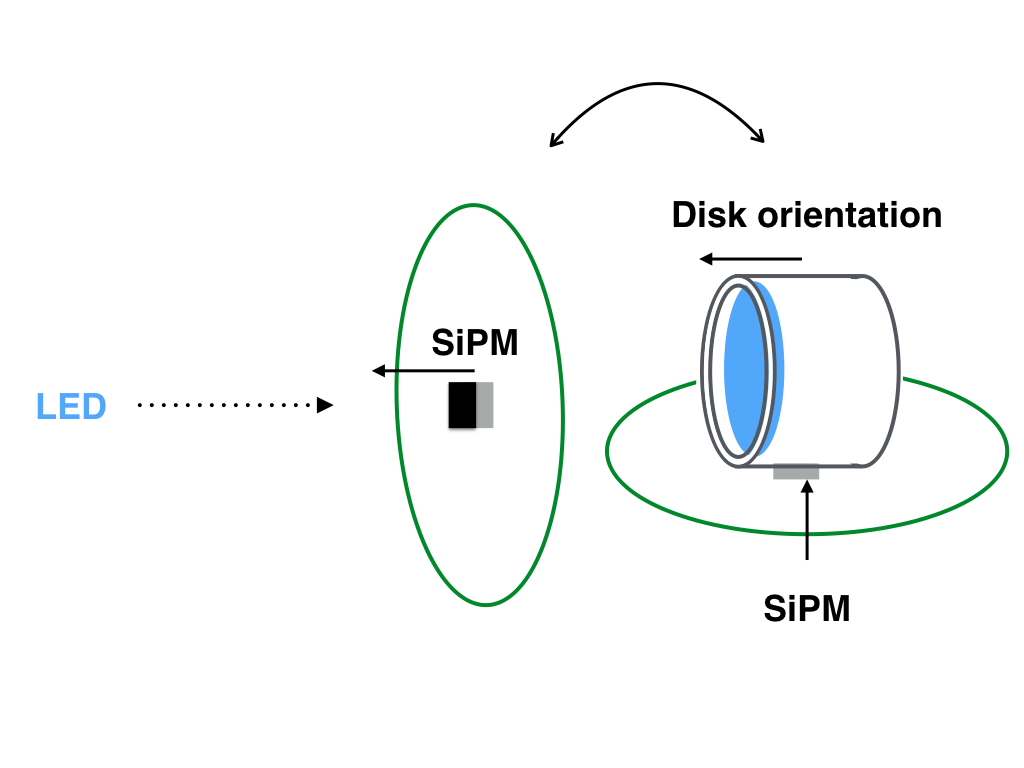}}
  \caption{Measurement overview.}
  \label{fig:LabMethod}
\end{figure}

The laboratory set-up (see Figure \ref{fig:scheme}) consisted of a light-tight black box containing a fast-response LED (NSHU590, peak output at 375nm) located behind an optical diffuser plate at one end with the light-sensor located 60 cm away. The LED driver is capable of generating LED flashes with widths on the order of nanoseconds (relevant to mimic the Cherenkov flashes from EAS), and is triggered via an external pulse generator. 

This detector setup can alternate between a "naked" SiPM (i.e. a SiPM with no scintillating disk) and the Light-Trap (see Figure \ref{fig:orient}). Data acquisition (full pulse trace or peak amplitude) was undertaken via a LabView program running on a laptop connected to a Rohde and Schwarz RTO 1024 2-GHz oscilloscope. Sets of measurements were taken with the LED flashing either the naked SiPM or the Light-Trap setup. The output signals for both sets of measurements were plotted into histograms and the mean signal amplitude for each device estimated via a Gaussian fit. A simple ratio of both the mean naked-SiPM signal and the Light-Trap output signal allowed for an estimation of the "boost" factor achieved by the additional use of the scintillating PMMA disk. The ratio of this boost factor with that expected by the simple geometric consideration (i.e. the increase in area between the 9 mm$^{2}$ KETEK SiPM and 176.71 mm$^{2}$ disk, a factor 19.63) gives the efficiency of the Light-Trap device. 

In terms of monitoring, there was no active temperature control within the box, however the laboratory temperature was controlled and monitored. Located next to the detector setup was an ET 9116A photomultiplier tube (PMT) - used as a brightness check for the LED.   

\subsection{Preliminary Laboratory Results}

\begin{figure}[htp]
  \centering
  \includegraphics[width=0.6\textwidth]{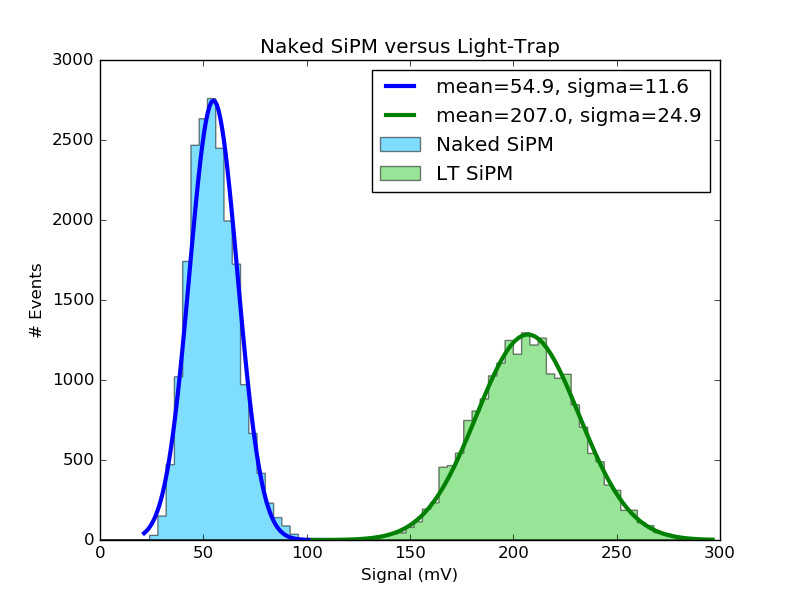}
  \caption{Distribution of signal peak-amplitudes from the uncovered SiPM (\emph{blue}) and Light-Trap (\emph{green}) when flashed by a 375 nm LED.}
  \label{fig:LTLab}
\end{figure}

Figure \ref{fig:LTLab} shows distributions of SiPM signal-amplitudes for the naked SiPM (blue histogram) and the Light-Trap system utilising the EJ-299-15 disk (green histogram). The average ratio of the signal means gives a boost factor of $\sim$3.8, corresponding to a Light-Trap efficiency of $\sim$20\%, the DSB1 disk also gave similar results. 

\section{Simulations}

\begin{figure}[htbp]
   \centering
   \includegraphics[width=0.4\textwidth]{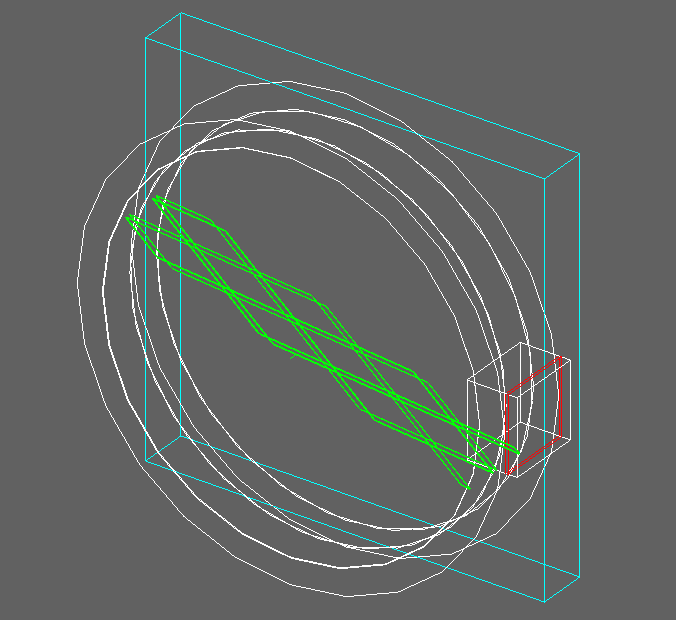} 
   \caption{Geant4 simulation of the Light-Trap. The SiPM is red, mirrors are cyan, and the green line is a TIR photon path within the disk.}
   \label{fig:LTSim}
\end{figure}

To help understand the performance and characteristics of the Light-Trap, a simple simulation of the system was created using the Geant4 software simulation package (version 4.10.01-p02)\cite{b}. A 15mm disk (optically perfect) with the absorption and emission properties of the EJ-299-15 material provided by Eljen was simulated. A coupling material (PMMA, with no dopant) was also added to optically couple a sensitive detector to the disk (i.e the SiPM, but the properties of the KETEK device have not been included in these simulations). Finally, two mirrors with 99\% reflectivity were placed at the bottom and sides of the disk, while leaving a 20 micron air gap. The simulated set-up can be seen in Figure \ref{fig:LTSim}.

Photons of a fixed energy (375 nm) were sent towards the open face of the disk, and the number of "hits" at the sensitive detector were then counted. The ratio between the number of photons sent to the disk versus those detected (times the quantum yield of the dopant, in the case of EJ-299-15 $\sim$84\%) gives an estimation of the overall "boost" factor, and hence the efficiency.

After 30,000 375nm photons were fired towards the disk (normal to the disk surface and over the full area), the percent efficiency was calculated to be 31.0\%. 

\section{Discussion}

When comparing the results of the laboratory measurements versus expectations from the Geant4 simulations, it is apparent that a discrepancy exists. The efficiency of the Light-Trap setup measured in the laboratory is $\sim$75\% of that expected from simulations. Several reasons for this mismatch have been considered. 

Firstly, the Light-Trap setup in the laboratory is simplistic, for example maintaining a small air gap around the disk is impossible and thus it is to be expected that this would have an impact on the TIR efficiency. While the silicone sheet used to optically couple the PMMA disk and SiPM is fully transmissive at the relevant wavelengths, it is attached to the disk via mechanical pressure, if any air-gaps exist at this interface it is likely to lead to a reduction in the number of photons reaching the SiPM. Furthermore, the 3M\textsuperscript{\textregistered} foil utilized in the setup may also suffer from losses when considering the whole spectral and incident angle range of photons. 
Finally, and perhaps most importantly, the optical surfaces of the PMMA disks provided by Eljen were studied using both a high-power microscope and through optical-bench testing in the laboratory to calculate scattering effects due to surface non-uniformities. The results of these studies revealed that the surfaces of the PMMA disks were not sufficiently smooth to provide efficient TIR (discussions over the optical polishing procedure undertaken by Eljen are ongoing to rectify this situation). 

The next step in the development of the Light-Trap is to receive new custom-doped PMMA disks from Eljen, however this time manufactured using a different industrial process to ensure the necessary optical-surface quality. Once these disks arrive at IFAE their optical performance will be evaluated before the proof-of-concept laboratory measurements are repeated. 

Furthermore, other approaches to the Light-Trap design and setup are being investigated, including the use of a Teflon disk-holder to avoid the need for reflective foil, a new design to decrease the air gap between the disk and holder to $\sim$20um (by using 3 small holding "fingers"), whether an alternative dye with a higher quantum yield could be used etc.

Studies of the PDE of the Light-Trap device and off-axis performance will also need to be conducted, before the development of a prototype Light-Trap pixel (LT-pixel) with a more advanced design. Ultimately, in collaboration with MPI-Munich, a cluster of seven LT-pixels is expected to be constructed and installed on one of the cameras of the MAGIC IACT experiment in La Palma, Spain \cite{magic1, magic2} for field-testing. 

\section{Conclusions}

A novel method to build relatively low-cost SiPM-based pixels utilising a disk of wavelength-shifting material has been proposed, and a proof-of-concept device has been constructed and tested in the lab. A Geant4 simulation of the device has also been created and used for performance predictions.

Initial measurements have shown that the Light-Trap concept works to first approximation, but that there are several steps to be undertaken to greatly enhance the performance of such a device. 

\acknowledgments

This project is supported by a Marie Sklodowska-Curie individual European fellowship (EU project 660138 --- Light-Trap) and Centro de Excelencia Severo Ochoa grant SEV-2012-0234. The authors also gratefully acknowledge the support of the engineering and technical staff at IFAE as well as the fruitful discussions undertaken with Eljen representatives.

\section*{Statement of Provenance}

This is an author-created, un-copyedited version of an article accepted for publication/published in the Journal of Instrumentation. IOP Publishing Ltd is not responsible for any errors or omissions in this version of the manuscript or any version derived from it. The Version of Record is available online at 10.1088/1748-0221/11/11/C11007

\end{document}